# An Energy-Efficient Edge Coprocessor for Neural Rendering with Explicit Data Reuse Strategies

Binzhe Yuan, Xiangyu Zhang, Zeyu Zheng, Yuefeng Zhang, Haochuan Wan, Zhechen Yuan, Junsheng Chen, Yunxiang He, Junran Ding, Xiaoming Zhang, Chaolin Rao, Wenyan Su, Pingqiang Zhou, Jingyi Yu and Xin Lou

*Abstract*—**Neural radiance fields (NeRF) have transformed 3D reconstruction and rendering, facilitating photorealistic image synthesis from sparse viewpoints. This work introduces an explicit data reuse neural rendering (EDR-NR) architecture, which reduces frequent external memory accesses (EMAs) and cache misses by exploiting the spatial locality from three phases, including rays, ray packets (RPs), and samples. The EDR-NR architecture features a four-stage scheduler that clusters rays on the basis of Z-order, prioritize lagging rays when ray divergence happens, reorders RPs based on spatial proximity, and issues samples out-of-orderly (OoO) according to the availability of on-chip feature data. In addition, a four-tier hierarchical RP marching (HRM) technique is integrated with an axis-aligned bounding box (AABB) to facilitate spatial skipping (SS), reducing redundant computations and improving throughput. Moreover, a balanced allocation strategy for feature storage is proposed to mitigate SRAM bank conflicts. Fabricated using a 40 nm process with a die area of 10.5 mm², the EDR-NR chip demonstrates a 2.41× enhancement in normalized energy efficiency, a 1.21× improvement in normalized area efficiency, a 1.20× increase in normalized throughput, and a 53.42% reduction in on-chip SRAM consumption compared to state-of-the-art accelerators.**

## I. INTRODUCTION

Neural radiance fields (NeRF) have facilitated substantial advancements in photorealistic scene reconstruction [1]. While 3D Gaussian Splatting (3DGS) enables faster training and interactive rendering on high-end GPUs, NeRF offers a smaller memory and bandwidth footprint, making it more viable for edge and resource-limited platforms [2].

Fig. 1 depicts the neural rendering pipeline in hash-based NeRF [3]. Once the scene geometry and texture are generated, images from arbitrary viewpoints can be synthesized through rendering. As the demand for NeRF-based applications continues to grow [4][5], integrating NeRF into edge devices has become increasingly critical. Extensive research has aimed to enhance NeRF processing speed and energy efficiency while preserving output quality [6]-[21].

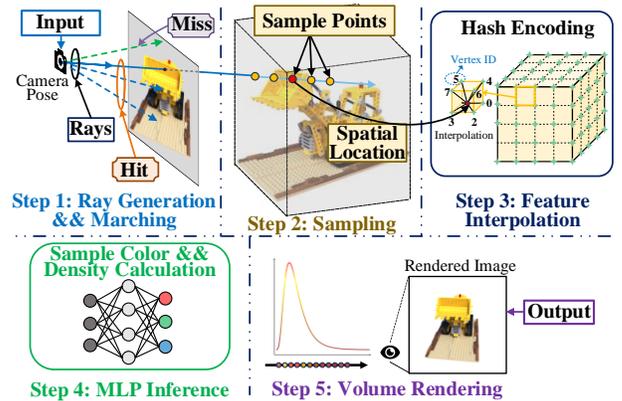

Fig. 1. The neural rendering pipeline in hash-based NeRF.

In contrast to GPUs, edge devices are often constrained by limited on-chip memory, leading to frequent external memory accesses (EMAs) and cache misses. Redundant EMAs increases energy consumption and degrade bandwidth efficiency. For NeRF, this issue deteriorates from the ray marching data flow, where sequential sampling along the same ray necessitates frequent voxel data transfers.

Many NeRF accelerators leverage spatial locality to mitigate excessive EMAs [6][7][8][14][17][20]. Adjacent rays are clustered to enhance data reuse. Most of these architectures employ row-order ray scanning, which is a straightforward approach that aligns with the pixel readout sequence of image sensors [22]. The row-order scanning simplifies implementation, but fails to fully harness spatial locality. This constraint stems from row-order scanning preference for grouping rays within the same row rather than accounting for spatial adjacency across both row and column dimensions. Recent studies [6][7] mitigate this constraint by organizing rays

 This work was partially supported by the Shanghai Rising-Star Program (Sailing Project) 23YF1427300. *(Co-first author: Xiangyu Zhang. Corresponding author: Xin Lou.)*

Binzhe Yuan, Xiangyu Zhang, Zeyu Zheng, Yuefeng Zhang, Haochuan Wan, Zhechen Yuan, Junshen Chen, Yunxiang He, Junran Ding, Wenyan Su, Pingqiang Zhou, Jingyi Yu and Xin Lou, are with the School of Information Science and Technology, Shanghaitech Univeristy, Shanghai, China.

Xiaoming Zhang, Chaolin Rao, are with the GGU Technology Co., Ltd, China.

into convolution-kernel-like patches, wherein row-order scanning is applied. A further challenge associated with ray grouping is ray divergence, which happens when rays within the same group traverse distinct voxels. The ray divergence diminishes ray-level parallelism and data reuse, remaining an insufficiently explored challenge in current literature.

Spatial skipping (SS) [23] has been investigated to enhance throughput by bypassing the sampling of empty regions [6][8][9][21]. Many implementations predominantly depend on voxel-based techniques yet do not explicitly optimize traversal path lengths. As a result, even in regions with a high likelihood of being empty, the occupancy grid bitmap of the corresponding voxels must still be retrieved [6][8]. Moreover, noise may lead to the misclassification of voxels as occupied, further diminishing the effectiveness of SS. According to [8], empty voxels comprise approximately 95–98% of the space. In addition, the skipping of empty voxels accounts for 29% of rendering cycles and induces pipeline bubbles.

Irregular memory access during hash table (HT) lookups induces SRAM bank conflicts. Many efforts have been put to address SRAM bank conflicts [6][8][9]. Reference [6] introduces an attention-based hybrid interpolation unit (AHIU), which selectively omits vertices that contribute minimally to interpolation. Reference [9] presents the vertex-interleaved mapping (VIM) technique, which facilitates one-time access to all vertex features linked to a voxel. Since vertex features are accessed across multiple adjacent voxels, redundant storage of these features leads to increased SRAM consumption.

Motivated by application demands, this paper introduces an explicit data reuse neural rendering (EDR-NR) architecture, that reduces EMAs overhead, expedites SS, and maintains rendering fidelity. The main contributions are as follows.

1) A four-stage scheduler designed to mitigate redundant EMAs by leveraging spatial locality from three phases. The four-stage scheduler integrates Z-order ray scanning, lag-first ray marching, ray packet (RP) clustering, and out-of-order (OoO) sample issuance, collectively enhancing on-chip data reuse and alleviating parallelism degradation caused by ray divergence.

2) A four-tier hierarchical RPs marching (HRM) methodology, integrated with an axis-aligned bounding box (AABB), shortens traversal paths and reduces redundant occupancy grid checks.

3) A balanced bank allocation strategy for the feature cache that mitigates bank conflicts, and eliminates redundant vertex feature storage.

The remainder of this paper is organized as follows. Section II provides an overview of the operational theory and a literature review. Section III discusses the optimization techniques. Section IV details the main components of the architecture. Section V presents the implementation results and a comparative analysis. Finally, Section VI concludes the paper.

## II. BACKGROUND AND RELATED WORK

### A. Instant Neural Graphics Primitives

Instant neural graphics primitives (Instant-NGP) is a NeRF variant that achieves high-quality rendering with remarkable efficiency [3]. The rendering pipeline consists of five stages as shown in Fig. 1. Firstly, rays are generated based on camera parameters and viewing direction. Secondly, samples are extracted along the ray. Thirdly, sample coordinates are mapped into high-dimensional feature vectors through interpolation. To achieve this, the voxel containing the sample is identified, and the indices of the eight vertices for voxel serve as keys in a hash function to compute feature addresses. The corresponding eight feature vectors are then retrieved from the HT and trilinearly interpolated into a single feature vector $f_{sam}$. Fourthly, $f_{sam}$ and the ray direction $d$ after frequency encoding go through a multilayer perceptron (MLP) to determine the sample color $c_i$ and density $\sigma_i$. Finally, the pixel color $C(r)$ for ray $r$ is computed by integrating the $N$ samples along the ray:

$$C(r) = \sum_{i=1}^{N} T_i \big(1 - exp(-\sigma_i \delta_i)\big) c_i \qquad (1)$$

where $\delta_i$ represents the distance between consecutive samples $sam_{i+1}$ and $sam_i$, and the transmittance $T_i$ is given by:

$$T_i = exp\big(-\sum_{j=1}^{i-1} \sigma_j \delta_j\big) \qquad (2)$$

### B. Related Work

Many studies have leveraged spatial locality to reduce EMAs and computational overhead via data reuse. The majority utilizes row-order scanning during ray generation and exploits ray-level parallelism. For instance, reference [6] proposes segmented hashing with spatial pruning (SHSP), resulting in a 66% reduction in EMAs. Furthermore, reference [6] clusters multiple rays into patches, and only stores patch addresses, thereby reducing memory usage by 88.3%. Reference [8] presents a voxel-centric data flow (VCDF), in which all samples within a voxel are generated prior to projection onto the image plane, leading to an 88.7% reduction in EMAs. Reference [20] processes samples sharing the same block identification (ID) within a processing-in-memory group (PIMG) and utilizes inter-patch block similarity, achieving a 75.6% block reuse rate. Reference [17] utilizes radiance proximity across rays from adjacent camera views through sparse radiance warping (SPARW), reducing radiance computations by up to 88%. Reference [14] presents utilization-driven memory replacement (UDMR), which incorporates a four-level pseudo-least-recently-used (P-LRU) strategy, reducing memory overhead by 94.6%, EMAs by 77.2%, and total computations by 90.9%. Reference [7] leverages temporal similarities across frames to reduce the number of pixels requiring rendering.

Recent advancements have significantly enhanced SS and improved throughput by bypassing sampling in empty voxels. For instance, [8] presents decoupled spatial skipping (DSS) and interleaved sampling (IS), which enhance sampling efficiency by 3.20× and improve rendering throughput by 2.41×. Reference [9] proposes a hierarchical empty space skipping (HESS) scheme, whereas reference [21] exploits occupancy grid sparsity to compute scene geometry directly. In addition, reference [6] removes redundant HT segments in empty space, leading to a 5.50× bitmap compression, and utilizes a hierarchical bitmap, resulting in a 3.67× increase in ray casting speed.

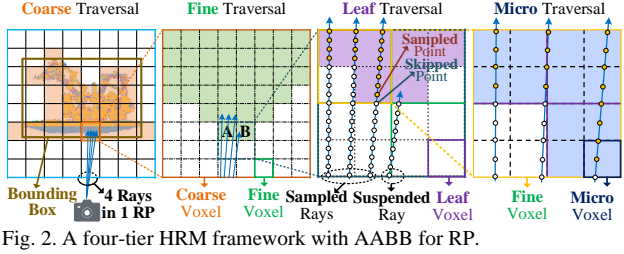

Fig. 2. A four-tier HRM framework with AABB for RP.

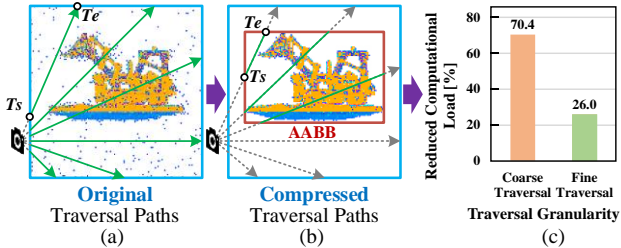

Fig. 3. (a) Original traversal paths, where $Ts$ and $Te$ denote the start and end points, respectively. (b) Compressed traversal paths using AABB intersection tests. (c) Computational load reduction achieved by introducing the AABB during coarse and fine stages.

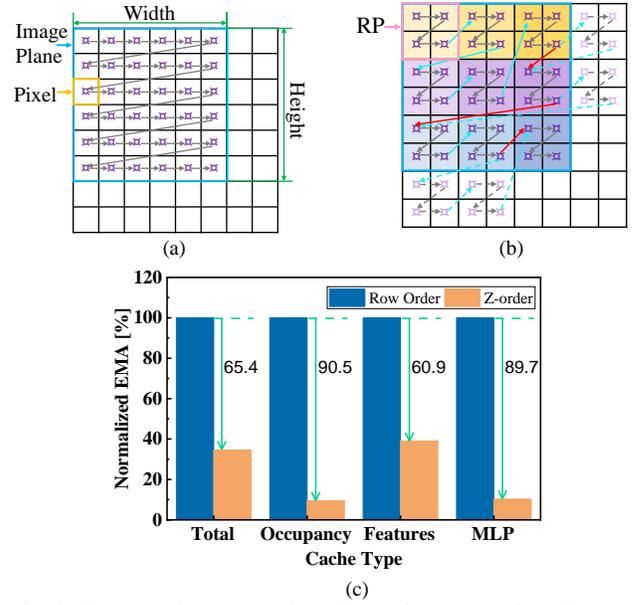

Fig. 4. (a) Row-order ray scanning. (b) Z-order ray scanning. (c) EMAs reduction by introducing Z-order.

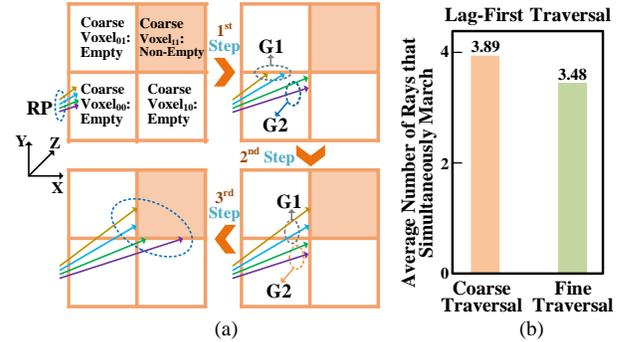

Fig. 5. (a) Illustration of lag-first traversal among coarse voxels. G1 and G2 denote Group 1 and Group 2, respectively. (b) The average number of rays that simultaneously march during coarse and fine traversal.

Current studies have made substantial progress in mitigating bank conflicts. Reference [6] introduces the AHIU to optimize heterogeneous memory access patterns, reducing power consumption by 56.4%. Reference [8] utilizes a base-Δ interpolation algorithm to optimize scattered memory accesses, enhancing the equivalent on-chip memory bandwidth by 2.38×. Reference [9] presents the VIM technique, distributing eight vertex features across eight banks for parallel fetching.

## III. HARDWARE-ORIENTED OPTIMIZATION TECHNIQUES

### A. HRM and AABB for SS

SS mitigates redundant computations by bypassing sampling in empty regions. However, the occupancy grid bitmap of all voxels must still be examined. To further refine SS while enhancing parallelism, this work proposes a four-tier HRM framework (Fig. 2) that integrates an AABB (Fig. 3) for RPs. The HRM framework comprises four tiers: coarse, fine, leaf, and micro traversal.

Unlike HESS [9], which is tailored for single-ray processing, the HRM framework operates on RPs encompassing multiple rays to increase parallelesim. A critical challenge for HRM is ray divergence within the RP, wherein rays propagate into distinct voxels. For example, as shown in Fig. 2, three rays enter voxel A while one enters voxel B during fine traversal. The HRM framework selectively advances rays within the same voxel while temporarily deferring others. A comprehensive analysis of ray divergence handling is presented in Section III-B.

Furthermore, the HRM framework incorporates an AABB module that utilizes a Gaussian filter to mitigate noise in occupancy grids, thereby constraining occupancy grid bitmap checks to voxels within the AABB. Consequently, traversal paths are compressed, leading to a 70.4% reduction in computational load during coarse traversal and a 26.0% reduction during fine traversal. The architectures of HRM and AABB are elaborated in Section IV-A.

### B. Spatial-Locality Oriented Scheduler for Data Reuse

Instant-NGP primarily relies on three categories of data requiring on-chip storage: the occupancy grid (micro grid) bitmap for SS, feature vectors, and MLP weights. Fully storing this dataset on-chip is infeasible. Instead, off-chip DRAM functions as external storage, facilitating selective retrieval into on-chip SRAM. Since EMAs introduce energy and latency overhead, this work proposes a four-stage scheduler that reduces repeated EMAs for the same data block by enhancing on-chip data reuse. Rays and RPs exhibiting high spatial locality are clustered, and samples are issued in an OoO manner based on available on-chip features.

The first-stage scheduler clusters four adjacent Z-order rays (Fig. 4(b)) into RPs, facilitating parallel traversal and data reuse. Compared to rays clustered via row-order scanning (Fig. 4(a)), Z-order rays within RPs demonstrate greater traversal path similarity, thereby increasing data access overlap to the occupancy grid, feature vectors, and MLP weights. The first-stage scheduler reduces EMAs by 65.4%. Architectural details

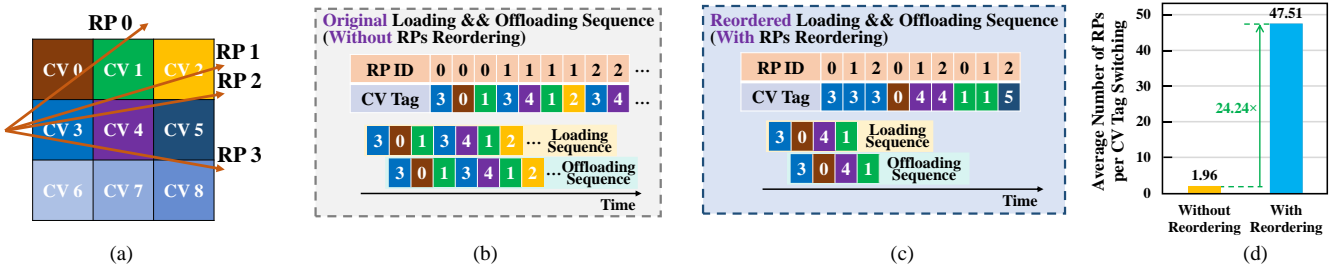

Fig. 6. (a) Coarse traversal of RPs. (b) EMA inefficiency without RPs reordering, where data from CV3, CV4, and CV1 must be repeatedly loaded and offloaded. (c) RPs reordering based on CV tags reduces EMA inefficiency. (d) The average number of RPs per CV tag switching is improved by 24.24×.

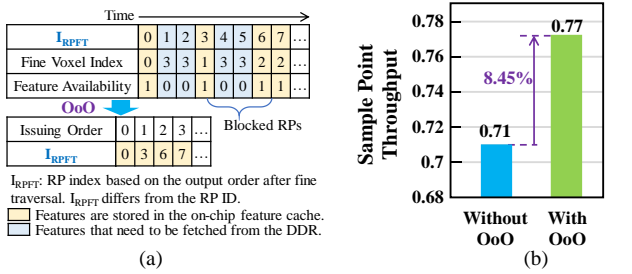

Fig. 7. (a) The OoO technique prioritizes issuing samples whose features are already on-chip. (b) Improvement in sample point throughput.

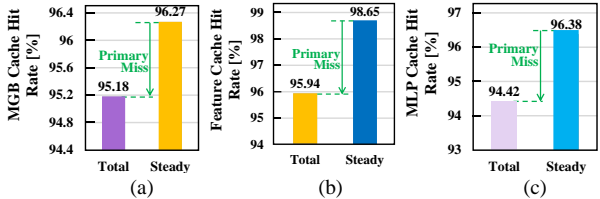

Fig. 8. Cache hit rates of (a) micro grid bitmap (MGB), (b) feature vectors, and (c) MLP weights.

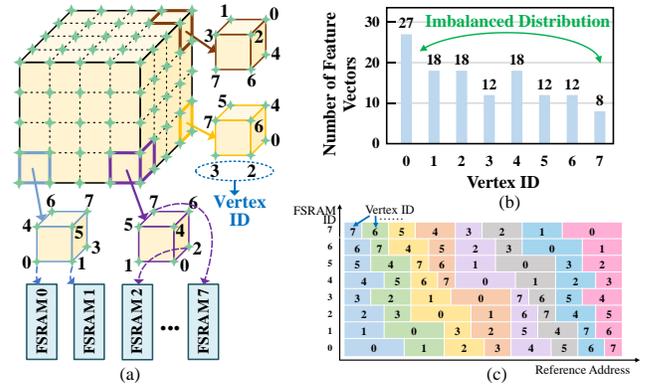

Fig. 9. (a) Storage of eight feature vectors in eight FSRAMs based on their vertex IDs. (b) The distribution of feature vectors for each vertex ID within a fine voxel is imbalanced. (c) Feature vector rearrangement within fine voxels to balance FSRAM depth. Each color represents the feature vectors of a single fine voxel.

of Z-order ray generation are provided in Section IV-B.

Ray divergence poses a critical challenge in RP-based marching, as rays within the same RP may enter distinct voxels (Fig. 2), diminishing ray-level parallelism and increasing repetitive EMAs. To mitigate throughput degradation caused by ray divergence, the second-stage scheduler employs a lag-first approach. Upon divergence, lagging rays are selectively advanced to promote convergence, as illustrated in Fig. 5(a). This technique attains an average of 3.89 and 3.48 simultaneous rays per RP during coarse and fine traversal, respectively. Architectural details of the lag-first approach are presented in Section IV-C.

The third-stage scheduler arranges RPs based on their coarse voxel (CV) tags (Fig. 6). On one hand, during RP marching, RPs traverse multiple CVs, and different RPs may access the same CV at different times, resulting in redundant EMAs. On the other hand, at CV boundaries, repeated loading and eviction of identical CV data further contribute to redundant EMAs. By reordering RPs based on their CV tags, these inefficiencies are alleviated, increasing the average number of RPs processed per CV tag switch by a factor of 24.24×. Architectural details of RP reordering are provided in Section IV-D.

The first three schedulers optimize data reuse during HRM, whereas the fourth scheduler targets frequent EMAs resulting from the unpredictable nature of feature accesses during interpolation. Given that each CV encompasses $8^3$ fine voxels,

RPs within the same CV may be distributed across distinct fine, leaf, and micro voxels. The fourth scheduler integrates a cache-aware sample issuance mechanism that monitors feature-cache hits and misses. Upon a feature-cache miss, subsequent RPs that register feature-cache hits are prioritized, enabling OoO execution (Fig. 7). The scheduler fetches the missed features while the RPs that register feature-cache hits are interpolating the features, hiding the EMA delay caused by cache misses. The OoO sample issuance enhances on-chip feature reuse, and reduces the processing time, yielding an 8.45% sample point throughput improvement. Architectural details of OoO sample issuance are presented in Section IV-E.

Together, these four schedulers improve cache hit rates, as illustrated in Fig. 8. Although primary cache misses are unavoidable during initial queries, the system attains steady-state cache hit rates of 96.27% (micro grid bitmap), 98.65% (feature vectors), and 96.38% (MLP weights), underscoring the effectiveness of the proposed scheduling techniques in reducing EMAs via enhanced data reuse.

### C. Balanced Bank Allocation for Feature Cache

The feature of a sample is tri-linearly interpolated from the eight vertexes features of the voxel that the sample belongs to. Similar to the VIM method in [9], this design stores vertex features using direct spatial indices as addresses. Each voxel vertex contains a feature vector, categorized into eight types according to vertex ID (Fig. 9(a)). The vertex classification enables direct storage alignment across eight memory banks, allowing all required features for a sample to be fetched in a single cycle.

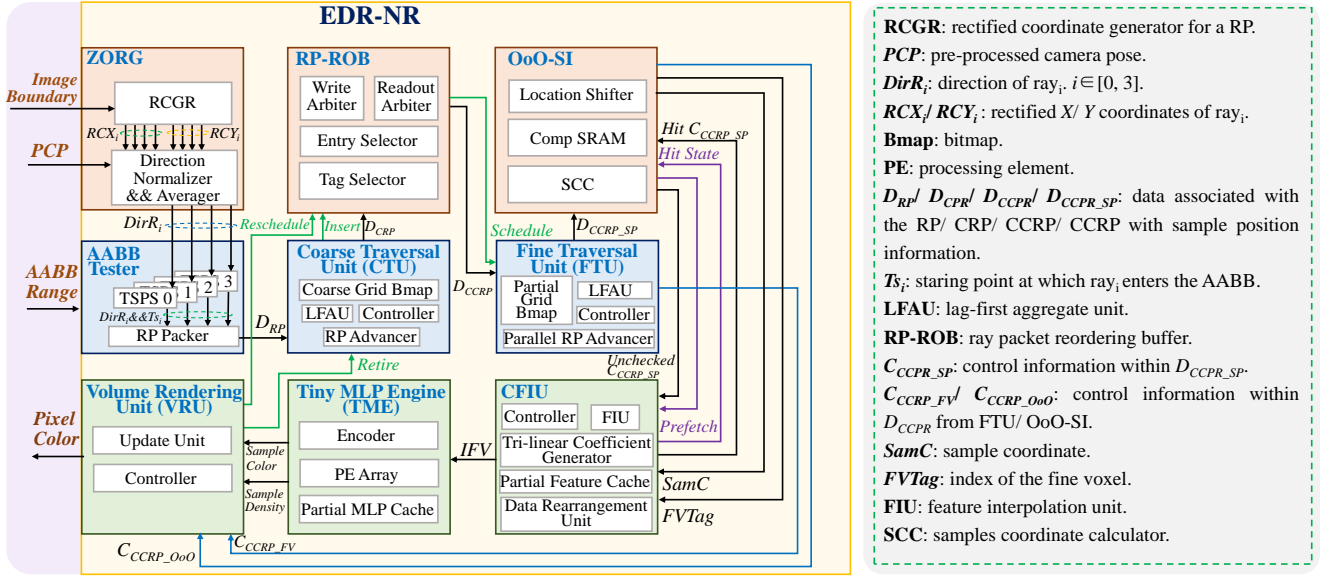

Fig. 10. Overall architecture of the proposed EDR-NR design.

Eliminating the HT results in storage redundancy. To mitigate this redundancy, feature vectors for an entire fine voxel are stored collectively instead of separately for each micro voxel in this work. The fine-voxel-centric storage reduces the number of feature vectors that need storage per fine voxel from 512 to 125. However, vertices shared among multiple fine voxels still necessitate redundant storage. Notably, because sample distributions are concentrated near object surfaces, accessed features are predominantly localized to these regions, diminishing the necessity for extensive storage across the entire scene [8].

A secondary challenge stems from the uneven distribution of feature vectors across the eight vertex types within a fine voxel (Fig. 9(b)). Directly mapping vertex IDs to feature SRAM (FSRAM) IDs results in non-uniform depth of FSRAM, increasing the difficulty in placement and routing [24]. To resolve the issue, this work introduces a bank-vertex ID mapping strategy that equalizes storage distribution across FSRAM banks (Fig. 9(c)). This strategy maintains uniform depth across banks. Architectural details of the balanced bank allocation strategy are presented in Section IV-F.

## IV. ARCHITECTURE AND DESIGN METHODOLOGY

### A. Overall Architecture and Operation Flow

Fig. 10 illustrates the nine key components of the EDR-NR architecture, categorized into three functional groups: (1) spatial locality optimization (red units) to enhance on-chip data reuse, (2) SS acceleration (blue units) to expedite valid sample localization, and (3) pixel computation (green units) to optimize resource efficiency.

The Z-order ray generator (ZORG) receives inputs from an external system, traverses the image plane in a Z-order pattern, and concurrently computes the directions of four rays. The AABB Tester (ABT) discards rays that fail to intersect objects, utilizing external AABB data. The remaining rays are assembled into RPs and processed via the HRM framework, which consists of the coarse traversal unit (CTU) and fine traversal unit (FTU). The traversal starting point searcher (TSPS) refines the RP entry point within the AABB using a binary search technique.

The CTU identifies the first non-empty CV within the AABB intersected by a RP. The RPs reordering buffer (RP-ROB) allocates an entry to store the candidate RP (CRP) from CTU. CRP is classified based on its CV tag, forming clustered CRP (CCRP), which is then forwarded to the FTU. The FTU detects the first non-empty fine voxel and generates samples as needed. Additionally, the FTU integrates leaf and micro traversal to efficiently reuse micro grid bitmap.

The CCRP, embedded with sample position information (CCRP_SP), is forwarded to the OoO sample issuer (OoO-SI). The OoO-SI instructs the conflict-free interpolation unit (CFIU) to prefetch the necessary features. The interpolated feature vectors (IFVs) are processed by the tiny MLP engine (TME) to compute sample color and density. The volume rendering unit (VRU) integrates sample colors along the ray to compute pixel values and transmittance.

The TME employs a spatial partitioning strategy similar to KiloNeRF, assigning each CV an independent fully connected network [25]. Each RP is assigned a unique pointer, serving as an address in the global RP buffer, and an identifier during the operation flow. The number of RPs under processing is decided by the capacity of the global RP buffer. A RP is considered fully processed once the transmittance of all the remaining samples drops below a predefined threshold or all its rays exit the AABB. After a RP completes rendering, four pixel values and their coordinates are output to reconstruct the new-view image.

As illustrated by the green path in Fig. 10, two feedback signals, retire and reschedule, regulate execution flow. The retire path is triggered under two conditions. (1) RP termination.

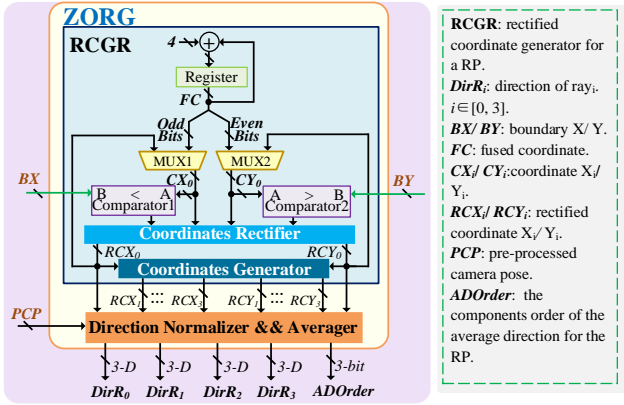

Fig. 11. ZORG architecture for parallel generation of four ray directions within a RP.

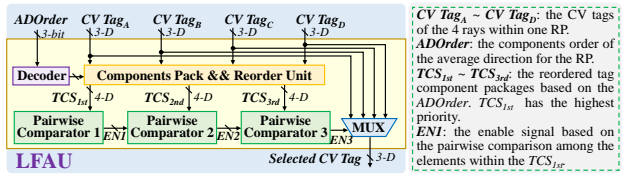

Fig. 12. LFAU architecture for aligning rays within a RP during coarse traversal.

The VRU detects that transmittance drops below a predefined threshold or that a RP exits the AABB during HRM. The RP pointer is released, and the RP-ROB deallocates a placeholder from the corresponding entry of RP. (2) CV transition. If a RP exits its current CV but remains within the AABB, the RP is redirected to the CTU via the retire path. Different from the RP termination condition, the placeholder is released whereas the RP pointer is kept. After the release, the RP behaves like a new RP from the ABT, and continues coarse traversal until the RP intersects a non-empty CV.

The reschedule path enables iterative processing due to multiple samples per ray and multiple rays per RP. Reschedule is triggered under fine voxel transition if the transmittance of current sample is still above the threshold. Continued sampling along the same ray or within the same RP is required. The ray length which is stored in the global RP buffer is updated.

### B. ZORG for Ray Generation

The ZORG microarchitecture (Fig. 11) employs a 22-bit counter to generate pixel coordinates along a Z-order curve. ZORG partitions the odd and even bits of the counter output ($FC$) into two 11-bit binary values, representing the X and Y coordinates ($CX$ and $CY$). Unlike row-order scanning traversal (Fig. 4(a)), which progresses in a predictable row-by-row manner, Z-order traversal (Fig. 4(b)) follows an irregular pattern, whereas improving the locality of rays within a RP. For images with non-power-of-two dimensions (e.g., 800×800), the Z-curve can extend beyond valid image boundaries, producing out-of-bounds coordinates before re-entering the valid region. To address this issue, the correlation between out-of-bounds positions and their subsequent re-entry points is analyzed. ZORG preemptively corrects coordinates before an out-of-bounds step occurs, ensuring that all rectified coordinates ($RCX$ and $RCY$) remain within valid bounds and reducing traversal stalls.

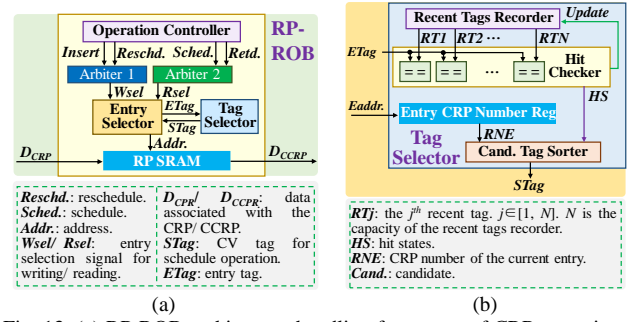

Fig. 13. (a) RP-ROB architecture handling four types of CRP operations. (b) Tag selector architecture determining the CV tag during tag switching.

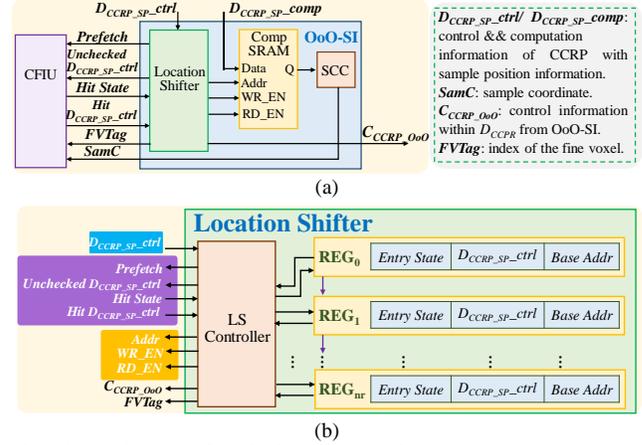

Fig. 14. Architecture of (a) OoO-SI and (b) LS.

### C. Lag-first Aggregate Unit (LFAU)

The LFAU (Fig. 12) orchestrates ray selection to address ray divergence in both CTU and FTU. The LFAU retrieves the voxel indices of four rays within the RP and sorts them along the X, Y, and Z axes. Pairwise comparisons along the three axes determine ray selection for traversal, producing three selection signals ($EN$). The rays with the Selected CV Tag that is determined by $EN3$ advance during the current iteration. After selecting prioritized rays, the CTU or FTU queries the corresponding coarse or fine bitmap to check voxel occupancy. If the queried bitmap designates a voxel as empty, the next voxel position is computed; otherwise, sampling proceeds.

### D. RP-ROB for Clustering RPs

The RP-ROB (Fig. 13(a)) organizes CRPs from the CTU by their CV tags, ensuring that CRPs with the same CV tag are stored in the same entry. The CCRPs are forwarded to the FTU. Because of spatial locality optimizations, CRP distribution across CVs is naturally imbalanced. A fixed allocation of entries per CV tag can lead to inefficient resource utilization or processing stalls. To address this, the RP-ROB dynamically assigns CV tags as entry labels, enabling CVs with more CRPs to occupy multiple entries when necessary.

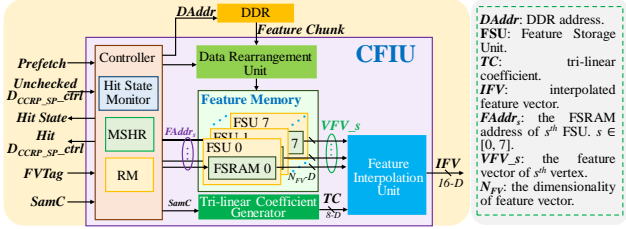

Fig. 15. Architecture of the CFIU.

<table>
<tr><td rowspan="3">Scene</td><td colspan="2">Chair</td><td colspan="2">Lego</td><td colspan="2">Mic</td><td colspan="2">Hotdog</td></tr>
<tr><td>Ground Truth</td><td>EDR-NR</td><td>Ground Truth</td><td>EDR-NR</td><td>Ground Truth</td><td>EDR-NR</td><td>Ground Truth</td><td>EDR-NR</td></tr>
</table>

| Output image | | | | | | | | |
|---|---|---|---|---|---|---|---|---|

| Avg. PSNR [dB] | 30.96 | | 33.75 | | 34.09 | | 35.64 | |

Fig. 16. Visual rendering results and PSNR of the EDR-NR chip.

The RP-ROB regulates four key signals to control the operation flow: insert, schedule, reschedule, and retire. Upon activation of the insert signal, an incoming CRP is mapped to an entry according to its CV tag, and the recent tags recorder (Fig. 13(b)) in the tag selector is updated following the least recently used (LRU) policy. If there are multiple entries having the same CV tag as the incoming tag, the entry selector gives priority to the most occupied entry. If the incoming CV tag is different from the tags of all the entries, and there exists empty entries, an empty entry is selected and labeled.

When the schedule signal is triggered, a CCRP is retrieved and sent to the FTU. If multiple entries correspond to the current CV tag, the entry selector selects the entry with the highest available capacity. Tag switching is necessary when all the CCRPs of current tag has retired. During tag switching, the tag selector gives preference to the tags in the recent tags recorder to improve CCRP locality. The entries with the hit tags are then sorted, and the entry with the least CCRPs is selected for schedule.

The reschedule and retire signals facilitate feedback mechanisms for managing iterative processing and rendering completion. A detailed discussion is provided in Section IV-A.

### E. OoO-SI for Sample Issuance

The OoO-SI (Fig. 14(a)) processes the CCRP containing sample positions from the FTU ($CCRP_{SP}$), instructs the CFIU to prefetch the required features, and out-of-orderly issues samples to the CFIU. Without the OoO-SI, stalls may arise when samples wait for feature retrieval from external memory. The OoO-SI utilizes an OoO queue (location shifter (LS)), prioritizing $CCRP_{SP}$ with available on-chip features, thereby reducing stalls caused by EMA delays. The LS (Fig. 14(b)) manages the queue, storing entry states, control information, and base addresses in registers. The base address specifies the storage location of the remaining $CCRP_{SP}$ data ($D_{CCRP\_SP}\_comp$) in the computation (Comp) SRAM. The sample coordinate calculator (SCC) generates the sample coordinate ($SamC$) which is forwarded to CFIU.

| | Specifications | | | |
|---|---|---|---|---|
| Technology | 40 nm 1P8M CMOS | | | |
| Area | 3.0 mm × 3.5 mm (10.5 mm²) | | | |
| SRAM [KB] | Micro Grid Cache | Feature Cache | MLP Cache | Others |
| | 8 | 256 | 8.25 | 43.2 |
| Supply Voltage [V] | 0.79-1.21 | | | |
| Maximum Frequency | 380 MHz | | | |
| Data Type | INT4, 8, 16, 20 | | | |
| Model Quantization | Feature: INT4, MLP: INT8 | | | |
| Operating Conditions | Power Consumption* [mW] | Rendering Speed* [FPS] | Energy Efficiency* [nJ/pixel] | |
| 50 MHz @ 0.79 V | 50.3 | 13.2 | 6.0 | |
| 150 MHz @ 0.98 V | 205.6 | 38.8 | 8.3 | |
| 250 MHz @ 1.06 V | 364.0 | 61.2 | 9.3 | |
| 350 MHz @ 1.16 V | 569.8 | 81.3 | 11.0 | |
| 380 MHz @ 1.21 V | 643.5 | 85.9 | 11.7 | |

*Estimated @ Synthetic NeRF Dataset.

### F. CFIU for Feature Interpolation

The CFIU (Fig. 15) handles feature interpolation, monitors feature availability status in the FSRAMs, and fetches required features from external memory in response to OoO-SI requests. To enhance memory access efficiency and avoid conflicts, feature vectors within a fine voxel are allocated across eight distinct FSRAMs based on vertex IDs.

The CFIU integrates miss status handling registers (MSHR) and a reservation monitor (RM). The MSHR aggregates duplicate miss requests to avoid repetitive EMAs, and RM prevents premature cache line replacement, ensuring data availability for ongoing computations. The tri-linear interpolation coefficients ($TCs$) are reordered to align FSRAM ID with the vertex ID.

## V. EVALUATION

Fig. 16 illustrates benchmark results for NeRF-based rendering, assessed with the Synthetic NeRF dataset [26]. Visual comparisons across multiple scenes between ground truth images and those rendered by the EDR-NR chip are showed. The EDR-NR chip attains a signal-to-noise ratio (PSNR) above 30 dB across these scenes, underscoring its rendering fidelity.

Table I and Fig. 17(a) provide an overview of the EDR-NR chip. Manufactured using 40 nm CMOS technology, the chip occupies a 10.5 mm² die area and operates with a supply voltage range of 0.79-1.21 V. At an 800×800 resolution, the chip reaches a peak rendering speed of 85.9 frames per second (FPS) while consuming 0.64 W at 380 MHz. The architecture incorporates 315.5 KB of on-chip SRAM, and supports custom floating-point operations for ray generation and HRM to enhance sample localization. Feature vectors utilize INT4 precision, whereas INT8 is employed for MLP weights. For real-time rendering at 38.8 FPS, the chip operates at 150 MHz and 0.98 V, consuming 0.21 W.

Fig. 17(b) depicts the demonstration system, which consists of a system board integrating the EDR-NR chip and a host PC

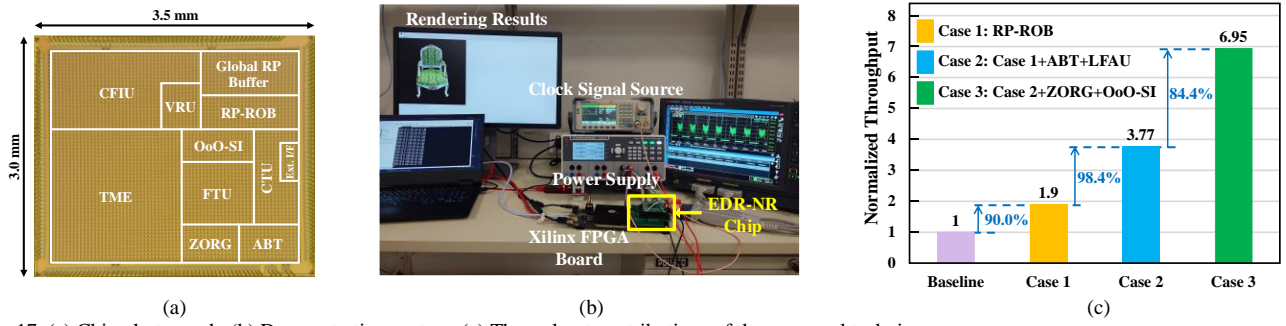

Fig. 17. (a) Chip photograph. (b) Demonstration system. (c) Throughput contributions of the proposed techniques.

TABLE II
PERFORMANCE COMPARISON OF THE EDR-NR CHIP WITH STATE-OF-THE-ART DESIGNS

|  | [27] | [28] | [7] | [20] | [6] | [9] | [8] | This Work |
|---|---|---|---|---|---|---|---|---|
| ASIC | × | × | ○ | ○ | ○ | × | × | ○ |
| Technology [nm] | 12 | 5 | 28 | 28 | 28 | 28 | 40 | 40 |
| Resolution | 800×800 | 800×800 | - | 800×800 | 800×800 | 800×800 | 800×800 | 800×800 |
| NeRF Model | Instant-NGP | Instant-NGP | Vanilla NeRF | Instant-NGP | Instant-NGP | Instant-NGP | Instant-NGP | Instant-NGP |
| Rendering Speed [FPS] | 2.9 | 75.6 | 110 | 30.6 | 73.5 | 120 | 131 | 85.9 |
| Area [mm²] | - | - | 20.25 | 5.07 | 20.25 | 15 | 19.38 | 10.5 |
| Voltage [V] | - | - | 0.6-0.95 | 1 | 0.68-0.9 | 0.9 | 0.9 | 0.79-1.21 |
| Max Frequency [MHz] | 1400 | 2500 | 250 | 200 | 200 | 300 | 400 | 380 |
| On-Chip Memory [KB] | - | - | 2015 | 360 | 2112 | 2,560 | 677.3 | 315.5 |
| Max Throughput [megapixel/second] | 1.86 | 48.38 | 1.44³⁾ | 19.58 | 47.04 | 76.8 | 83.84 | 54.98 |
| Normalized Throughput [megapixel/second/mm²] | - | - | 0.03 | 3.86 | 2.32 | 5.12 | 4.33 | 5.23 |
| Power [W] | 15²⁾ | 350²⁾ | 0.90 | 0.13 | 0.73 | 1.9 | 1.30 | 0.64⁵⁾ |
| Energy Efficiency [nJ/pixel] | - | - | 544.4³⁾ | 6.6 | 15.5⁴⁾ | 25 | 15.51 | 11.7⁵⁾ |
| Normalized Energy Efficiency [FPS/W] ¹⁾ | - | - | 36.92 | 78.77 | 27.29 | 24.46 | 55.58 | 134.22 |
| Normalized Area Efficiency [FPS/mm²] ¹⁾ | - | - | 1.864 | 2.07 | 1.253 | 2.74 | 6.76 | 8.18 |

1) Normalized to 40 nm technology and 1.21 V using the methodology outlined in [8] [29] [30]. $s$=Technology/40nm, $f \sim s$, $A \sim 1/s^2$ and $P \sim (1/s) \times (1.21/Voltage)^2$, where $f$, $A$ and $P$ denote frequency, area, and power, respectively. The normalized energy efficiency and normalized area efficiency are evaluated at the maximum rendering speed.
2) Thermal design power.
3) Not explicitly reported in the original paper, calculated based on the reported energy per sample and energy per pixel.
4) Estimated based on the reported FPS and an assumed resolution of 800×800.
5) Including power consumption from external memory access (measured with DDR3 SDRAM).

for display. The throughput breakdown in Fig. 17(c) indicates that the proposed architectural techniques yield a 6.95× increase in throughput, with ZORG and OoO-SI contributing the largest share (45.76%). Serving as the first processing unit in EDR-NR, ZORG generates Z-order rays, improving RP spatial locality compared to conventional row-order scanning. Additionally, the OoO-SI enhances feature cache hit rates. The feature cache occupies 81.14% of the total on-chip SRAM.

Table II presents a comparison of the EDR-NR chip with state-of-the-art accelerators, including NeRF ASICs [6][7][20], edge GPUs [27][28], and simulation-based implementations [8][9]. For a fair comparison, energy and area efficiency metrics are normalized according to the methodology in [8][29][30], which accounts for frequency, voltage, and technology differences.

As shown in Table II, the recent accelerator [20] achieves the highest energy efficiency while maintaining the lowest area consumption, delivering 30.6 FPS. Simulation-based implementations, including [8] and [9], as well as ASIC for vanilla NeRF [7], exceed 110 FPS. The EDR-NR chip achieves low on-chip memory usage while sustaining high normalized throughput, leading to a 1.21× increase in normalized area efficiency over [8]. Moreover, by effectively reducing EMAs, the EDR-NR chip attains a 2.41× improvement in normalized energy efficiency relative to [8].

## VI. CONCLUSIONS

This paper introduces the EDR-NR architecture, designed for rendering on edge devices. The EDR-NR chip improves energy and area efficiency through three key factors. First, the four-stage scheduler coordinates spatial-locality-aware ray packing, RP marching, and sample issuance, enhancing data reuse and increasing cache hit rates. Second, HRM, in conjunction with AABB, accelerates sample positioning, thereby increasing throughput. Third, balanced bank allocation reduces bank conflicts of feature cache, which accounts for 81.14% of the total on-chip SRAM. In summary, the EDR-NR chip effectively mitigates challenges in energy efficiency and resource consumption, establishing it as a viable solution for edge computing.

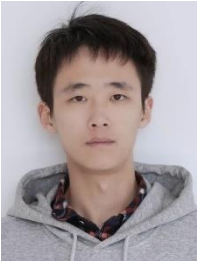
**Binzhe Yuan** received the B.Eng. degree in Electronic Information Engineering from ShanghaiTech University, Shanghai, China, in 2022. He is currently pursuing the M.Eng. degree in ShanghaiTech University, Shanghai, China. His current research interests include computer arithmetic and high-performance neural rendering accelerator design.

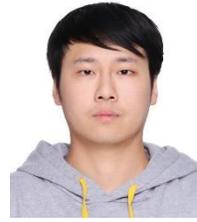
**Zhechen Yuan** (Student Member, IEEE) received the B.Eng. degree in Electronics Engineering from Shanghaitech University, China, in 2021. He is currently working toward the Ph.D. degree in Shanghaitech University. His research interests include neural rendering and energy-efficient VLSI design for computer graphics and deep learning.

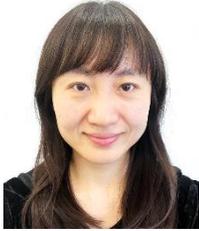
**XiangYu Zhang** (Member, IEEE) obtained her B.S. in Electronics and Electrical Science and Engineering from Tianjin University, Tianjin, China 2013. She continued her academic journey with an M.S. degree in Electrical Science and Engineering and a Ph.D. degree in System Cybernetics, both from Hiroshima University, Hiroshima, Japan, in 2016 and 2019, respectively. Presently, Zhang serves as an assistant researcher at ShanghaiTech University, Shanghai, China, where her research primarily focuses on the development of effective and efficient hardware accelerators.

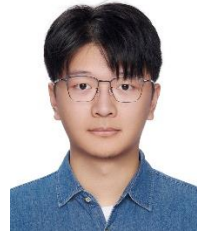
**Chen Junsheng** (Graduate Student, School of Information Science and Technology, ShanghaiTech University, Shanghai, China) received the B.Eng. degree in Electronic Information Engineering from ShanghaiTech University, Shanghai, China, in 2024, and is currently pursuing the M.S. degree in Electronic Science and Technology at ShanghaiTech University, Shanghai, China. His current research interests include neural rendering, hardware–software co-design, FPGA/ASIC accelerator design, and domain-specific hardware architectures.

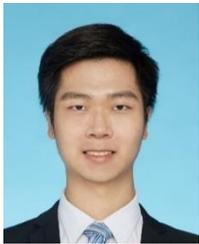
**Zeyu Zheng** received the B.Eng. degree in Computer Science from ShanghaiTech University, China, in 2022. He is currently pursuing the Master degree in Electronic Information. He is currently working for co-operate program of Shanghai GGUTech Co. Ltd. and ShanghaiTech University, focusing on test platform building and driver development for FPGAs and ASICs.

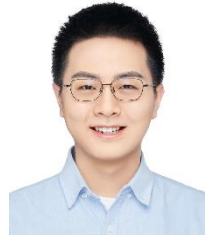
**Yunxiang He** is currently working toward the B.S. degree at ShanghaiTech Univeristy, Shanghai, China. His research interests include the architecture of custom accelerators based on vision, computer graphics, and deep learning.

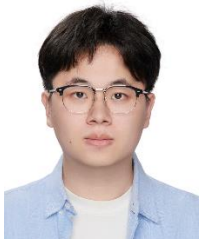
**Yuefeng Zhang** received the B.Eng. degree in Electronic Information Engineering from ShanghaiTech University, China, in 2024. He is currently working toward the D.Eng. degree in ShanghaiTech University. His research interests include neural rendering and VLSI design for computer graphics.

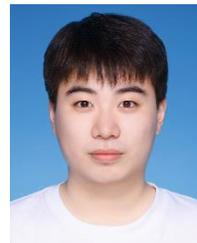
**Junran Ding** (Student Member, IEEE) received the B.Eng. degree in Electronic Information Engineering from ShanghaiTech University, Shanghai, China, in 2023. He is currently pursuing the M.Eng. degree in ShanghaiTech University, Shanghai, China. His current research interest includes hardware software co-design for neural rendering.

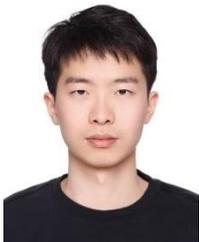
**Haochuan Wan** (Graduate Student Member, IEEE) received his B.Eng. degree from ShanghaiTech University, Shanghai, China, in 2021. He is currently pursuing a Ph.D. degree at the School of Information Science and Technology, ShanghaiTech University. His research interests include digital ASIC design and hardware acceleration for neural rendering algorithms.

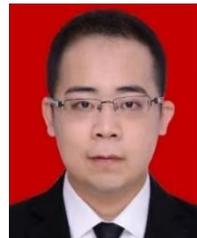
**Xiaoming Zhang** received the B.S. degree in electrical engineering from Jilin University, Changchun, China, in 2020, and the M.S. degree in electrical engineering from ShanghaiTech University, Shanghai, in 2023. He is currently with GGU Technology Company Ltd，Shanghai. His current research interests include neural rendering and VLSI design for artificial intelligence.

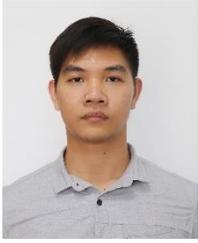

**Chaolin Rao** received the B.Eng. degree from the University of Electronic Science and Technology of China, Chengdu, China, in 2016, and the Ph.D. degree from the School of Information Science and Technology, ShanghaiTech University, Shanghai, China. He is currently with GGU Technology Company Ltd. His current research interests include neural rendering, computer architecture, and VLSI design for artificial intelligence.

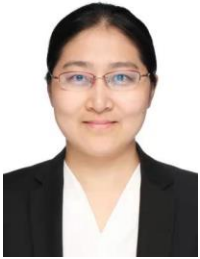

**Wenyan Su** received the B.Eng. degree in Information Engineering from East China University of Science and Technology, China, in 2007 and M.Sc. degree in Microelectronics from Fudan University, China, in 2010. She received the second M.Sc degree in System-on-Chip from Royal Institute of Technology(KTH), Sweden, in 2010 through the Fudan-KTH joint program.

Before joining ShanghaiTech University, Shanghai, China, she was the senior Engineer in LSI Technologies, Shanghai, China from 2010 to 2015, and then joined as the Senior Staff Engineer in IBM, Shanghai, China from 2015 to 2021. She did design service works using different advanced node technologies. As an Engineer in ShanghaiTech University now, her current research interests include high-performance and energy-efficient integrated circuits design.

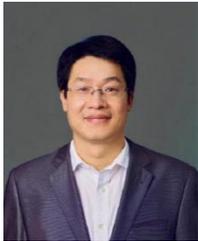

**Pingqiang Zhou** (Member, IEEE) received the B.E. degree from Nanjing University of Posts and Telecommunications, China, in 2005, the M.E. degree from Tsinghua University, Beijing, China, in 2007, and the Ph.D. degree from the University of Minnesota in 2012. He is currently a full professor with the School of Information Science and Technology at ShanghaiTech University, Shanghai, China. Prior to joining ShanghaiTech, he worked respectively at IBM T. J. Watson Research Center as a research intern in 2011, and the University of Minnesota as a postdoctoral researcher from 2012 to 2013. He was with the University of California, Berkeley as a visiting scholar in 2015. His current research interests include the computer-aided design of VLSI circuits, computer architecture, and hardware security. Prof. Zhou received the best paper nominations in ASP-DAC 2010 and CSTIC 2016. He has been serving on the technical program committees of many international conferences such as DAC, ICCAD, and ASP-DAC, and is an associate editor of the IEEE Transactions on Circuits and System II.

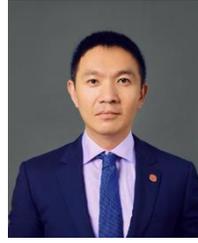

**Jingyi Yu** (Fellow, IEEE) received BS from Caltech in 2000 and Ph.D. from MIT in 2005. He is currently the Vice Provost at ShanghaiTech University. Before joining ShanghaiTech, he was a full professor in the Department of Computer and Information Sciences at University of Delaware. His research interests span a range of topics in computer vision and computer graphics, especially on computational photography and nonconventional optics and camera designs. He is a recipient of the NSF CAREER Award and the AFOSR YIP Award, and has served as an area chair of many international conferences including CVPR, ICCV, ECCV, IJCAI and NeurIPS. He was a program chair of CVPR 2021 and will be a program chair of ICCV 2025.

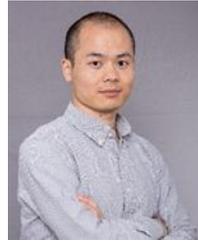

**Xin Lou** (Senior Member, IEEE) received the B.Eng. degree in Electronic Information Technology and Instrumentation from Zhejiang University (ZJU), China, in 2010 and M.Sc. degree in Systemon-Chip Design from Royal Institute of Technology (KTH), Sweden, in 2012 and PhD degree in Electrical and Electronic Engineering from Nanyang Technological University (NTU), Singapore, in 2016. Then he joined VIRTUS, IC Design Centre of Excellence at NTU as a research scientist. He is currently an Associate Professor with the School of Information Science and Technology, ShanghaiTech University, Shanghai, China. His research interests primarily focus on high-performance and energyefficient integrated circuits and systems for vision and graphics processing. Dr. Lou is an Associate Editor of IEEE Transactions on Circuits and Systems II: Express Briefs and a TPC member of the Circuits and Systems Society.